\documentclass[conference]{IEEEtran}
\IEEEoverridecommandlockouts

\usepackage{cite}
\usepackage{amsmath,amssymb,amsfonts}
\usepackage{graphicx}
\usepackage{textcomp}
\usepackage{xcolor}
\usepackage{float}
\usepackage{algorithm}
\usepackage{algpseudocode}
\usepackage{bm}
\usepackage{tabularx}
\usepackage{multirow}
\usepackage{booktabs}
\usepackage{stfloats}
\usepackage{ntheorem}
\usepackage{subfigure}

\makeatletter
\renewcommand{\fnum@algorithm}{}

\begin{document}
\title{
Dual-Mapping Sparse Vector Transmission for\\ Short Packet URLLC
\vspace{-0pt}
}
%\title{BEM Coefficients Interpolation based Channel Estimation for OFDM System over Sparse Doubly Selective Channel}
%\thanks{This work was supported in part by the National Natural Science Foundation of China under Grant 61901138, in part by the Natural Science Foundation of Guangdong Province under Grants 2018A030313344 and 2018A030313298, and in part by the Guangdong Science and Technology Planning Project 2018B030322004.}

%\author{
%Yanfeng Zhang$^{\star}$, Xu Zhu$^{\dagger}$, Yujie Liu$^{\#}$, Guan Yong Liang$^{\#}$, Hui Liang$^{\star}$, Ning Zhang$^{\sharp}$, and Xuemin Shen$^{\S}$\\
%$^{\star}$ School of Electrical Engineering \& Intelligentization, Dongguan University of Technology, Dongguan, China\\
%$^{\dagger}$ School of Electronic and Information Engineering, Harbin Institute of Technology, Shenzhen, China\\
%$^{\#}$ Continental-NTU Corporate Laboratory, Nanyang Technological University, Singapore\\
%$^{\sharp}$ Department of Electrical and Computer Engineering, University of Windsor, Windsor, Canada\\
%$^{\S}$ Department of Electrical and Computer Engineering, University of Waterloo, Waterloo, Canada\\
%Emails: yfzhang@ieee.org, xuzhu@ieee.org and sshen@uwaterloo.ca}

\author{
Yanfeng Zhang$^{\star}$, Xu Zhu$^{\dagger}$, Jinkai Zheng$^{\star}$, Weiwei Yang$^{\star}$, Xianhua Yu$^{\star}$, \\Haiyong Zeng$^{\#}$, Yujie Liu$^{\S}$, and Yong Liang Guan$^{\S}$  \\
$^{\star}$School of Electrical Engineering \& Intelligentization, Dongguan University of Technology, Dongguan, China\\
$^{\dagger}$School of Electronic and Information Engineering, Harbin Institute of Technology, Shenzhen, China\\
$^{\#}$School of Electronic and Information Engineering, Guangxi Normal University, Guilin, China\\
%$^{\S}$School of Engineering, Newcastle University, Newcastle upon Tyne, U.K.\\
%$^{\ddag}$ Frontier Research Center, Pengcheng Laboratory, Shenzhen, China\\
%$^{\S}$Department of ECE, The Hong Kong University of Science and Technology, Hong Kong, China\\
%$^{\S}$ School of Cyber Science and Engineering, Xi'an Jiaotong University, Xi'an, China\\
$^{\S}$Continental-NTU Corporate Laboratory, Nanyang Technological University, Singapore\\
Emails: yfzhang@ieee.org, xuzhu@ieee.org\vspace{-0pt}
\thanks{The work was supported in part by the Guangdong Basic and Applied Basic Research Foundation under Grant 2024A1515110033 and 2024A1515110036; in part by the Dongguan Strategic Scientist Teams Project under Grant 20231900700022; in part by the Natural Science Foundation of China under Grants 62171161 and 62301172; in part by the Guangdong Provincial Basic and Applied Basic Research Foundation under Grant 2022B1515120018; in part by the Challenge-Driven Program of the Ministry of Industry and Information Technology; and in part by the Shenzhen Science and Technology Program under Grants GXWD20220817133854003.  }\vspace{-0pt}
}

%\author{
%Yanfeng Zhang$^{\star}$, Xu Zhu$^{\dagger}$, Yufei Jiang$^{\dagger}$, Hui Liang$^{\star}$, Ning Zhang$^{\sharp}$, and Xuemin Shen$^{\S}$\\
%$^{\star}$ School of Electrical Engineering \& Intelligentization, Dongguan University of Technology, Dongguan, China\\
%$^{\dagger}$ School of Electronic and Information Engineering, Harbin Institute of Technology, Shenzhen, China\\
%%$^{\#}$ Continental-NTU Corporate Laboratory, Nanyang Technological University, Singapore\\
%$^{\sharp}$ Department of Electrical and Computer Engineering, University of Windsor, Windsor, N9B-3P4, Canada\\
%$^{\S}$ Department of Electrical and Computer Engineering, University of Waterloo, Waterloo, N2L-3G1, Canada\\
%Emails: yfzhang@ieee.org, xuzhu@ieee.org and sshen@uwaterloo.ca}

\maketitle

\begin{abstract}
Sparse vector coding (SVC) is a promising short-packet transmission method for ultra reliable low latency communication (URLLC) in next generation communication systems. In this paper, a dual-mapping SVC (DM-SVC) based short packet transmission scheme is proposed to further enhance the transmission performance of SVC. The core idea behind the proposed scheme lies in mapping the transmitted information bits onto sparse vectors via block and single-element sparse mappings. The block sparse mapping pattern is able to concentrate the transmit power in a small number of non-zero blocks thus improving the decoding accuracy, while the single-element sparse mapping pattern ensures that the code length does not increase dramatically with the number of transmitted information bits. At the receiver, a two-stage decoding algorithm is proposed to sequentially identify non-zero block indexes and single-element non-zero indexes. Extensive simulation results verify that proposed DM-SVC scheme outperforms the existing SVC schemes in terms of block error rate and spectral efficiency.
\end{abstract}

%\addtolength{\topmargin}{-0.49in}
%\addtolength{\topmargin}{-0.87cm}
\vspace{-0pt}
\section{Introduction}
Ultra reliable low latency communication (URLLC) is one of the service categories in next generation mobile communication systems, designed to support latency-sensitive applications such as autonomous driving, Internet of Things (IoT), and remote sensing control \cite{Kim20201,Jingxiaoye,YuxianhuaTWC}. The existing long data packet transmission paradigm is difficult to adapt to URLLC scenarios. On the one hand, a significant amount of resources needs to be allocated to signaling and error checking to ensure reliability, leading to high transmission latency and low spectral efficiency (SE). On the other hand, as the transmission block length decreases, the redundancy of existing low-density parity-check (LDPC) codes will decrease, and polar codes will not be able to fully polarize the channel, resulting in low reliability. Therefore, designing effective short-packet transmission methods is a critical issue in URLLC.

Recently, a short-packet transmission method, called sparse vector coding (SVC), has been proposed for \mbox{URLLC \cite{Ji2018}}. Unlike traditional methods, SVC maps the information bits into the non-zero indexes of a sparse vector, and then transmits the randomly spread sparse vector in time-frequency (TF) domain. The receiver only needs to identify the non-zero indexes of received signal through a simple energy detection algorithm to achieve decoding. Lots of studies have demonstrated that SVC has higher reliability compared to traditional channel coding techniques in short-packet transmission \cite{Ji2018,Kim20202,ZhangXuewan2022,YangLinjie2024}.

A number of SVC variants have been proposed to enhance the performance of original SVC scheme. 
In \cite{Kim20202}, an enhanced SVC (ESVC) scheme has been proposed, in which the transmitted information bits are mapped into both non-zero indexes and quadrature amplitude modulation (QAM) symbols. In \cite{ZhangXuewan2022}, the sparse superimposed codes (SSC) with multiple different constellations has been proposed to improve decoding performance. Moreover, the indexes of different constellation alphabets are utilized to carry additional information bits to further improve the block error rate (BLER) performance \mbox{in \cite{YangLinjie2024}.} Similar to SVC scheme, the sparse regression code (SPARC) is also a popular sparse vector transmission \mbox{method \cite{CRushTIT2021}.} The approximate message passing (AMP) algorithm has been proposed as a computationally-efficient decoder for SPARC \cite{CRushTIT2021}. Recently, a generalized SPARC (GSPARC) has been proposed in \cite{Sinha2024TCOM}. By constructing codewords through a sequential bit mapping strategy over multiple orthogonal sub-matrices, a block orthogonal sparse superposition (BOSS) code has been proposed in \cite{DhanTWC2023}. In our previous work, a block SVC scheme has been proposed by designing a block sparse mapping \mbox{pattern \cite{yfzhang2024BSVC}.} The concept of SVC has been extended to diverse communication systems, including multiple-input multiple-output (MIMO) systems \cite{ZhangRuoyu2021}, communication systems with low-resolution ADCs \cite{ZYF2024ICCC}, high-mobility communications \cite{ZhangYf2023}, semantic communications \cite{Zhanxunyang25}, massive grant-free access \cite{Luoyingzhe24}, URLLC with low storage overhead \cite{ZYFWCNC25} etc.

Most of the SVC works in \cite{Ji2018,Kim20202,ZhangXuewan2022,CRushTIT2021,Sinha2024TCOM,DhanTWC2023,yfzhang2024BSVC,YangLinjie2024,ZhangYf2023,
ZhangRuoyu2021,ZhangXuewan2021,ZYF2024ICCC} focuses on the design of sparse mapping patterns and the development of decoding algorithms. However, these SVC schemes only consider single-element sparse mapping pattern and have not yet explored the enhancement mechanism of multiple sparse mapping patterns on SVC encoding efficiency. More importantly, little work has been done to investigate how to utilize the priori information in terms of the distribution of non-zero elements in sparse vectors to design efficient decoding algorithms. Therefore, how to utilize multiple sparse mapping patterns to enhance the transmission performance of SVC remains an open topic.

In this paper, a dual-mapping SVC (DM-SVC) scheme is proposed for short-packet transmission in URLLC scenarios. At the transmitter, the transmitted information bits are mapped to a sparse vector by block and single-element sparse mapping. At the receiver, a two-stage decoding algorithm is proposed to sequentially identify the indexes of non-zero block and non-zero elements to achieve DM-SVC decoding. The main contributions of this paper are summarized as follows.

\begin{itemize}
\item To our best knowledge, this is the first work to introduce two sparse mapping patterns in SVC scheme to improve SE. Unlike the single-element random sparse mapping pattern used in conventional SVC schemes \cite{Ji2018,Kim20202,ZhangXuewan2022,CRushTIT2021,Sinha2024TCOM,DhanTWC2023,yfzhang2024BSVC,YangLinjie2024,ZhangYf2023,
ZhangRuoyu2021,ZhangXuewan2021,ZYF2024ICCC}, the proposed DM-SVC scheme uses two sparse mapping patterns to map information bits onto indexes of non-zero blocks and non-zero elements. Benefiting from the block sparse mapping pattern, the number of TF resources required for transmission is significantly reduced, which improves the SE of data transmission. In addition, the single-element sparse mapping pattern ensures that the code length does not increase dramatically with the number of transmitted information bits.

\item A two-stage algorithm is proposed for accurate decoding. By utilizing the block-structured priori information of non-zero elements in sparse vectors, the proposed decoding algorithm is able to sequentially identify the indexes of non-zero blocks and non-zero elements more accurately than the existing AMP \cite{CRushTIT2021} and multipath matching pursuit (MMP) \cite{Kwon2014} algorithms. Simulation results verify that the
DM-SVC scheme outperforms the existing \mbox{ESVC \cite{Kim20202},} SSC \cite{ZhangXuewan2022}, BOSS \cite{DhanTWC2023} and GSPARC \cite{Sinha2024TCOM} schemes in terms of BLER performance. 
\end{itemize}

%The rest of this paper is organized as follows. In Sections II and III, the DM-SVC encoding and decoding are described, respectively. In Section IV, the complexity of DM-SVC decoding algorithm is analyzed. Extensive simulation results are shown in Section V, while conclusions are drawn in \mbox{Section VI}.

%\emph{Notations:} Bold symbols represent vectors or matrices. ${( \cdot )^{\rm T}}$, ${( \cdot )^{\rm H}}$ and ${( \cdot )^{ \dag }}$ denote the transpose, conjugate transpose and matrix inversion, respectively. $|| \cdot |{|_p}$ denotes ${\ell _p}$ norm operation. $\left\lfloor  \cdot  \right\rfloor$ is the round-down operation. ${\tbinom{N}{K}}$ denotes the number of combinations of selecting $K$ items from $N$ items. 

%\addtolength{\topmargin}{-0.87cm}
\vspace{-0pt}
\section{Encoding of DM-SVC}
In this section, the concept of DM-SVC encoding is introduced, including the sparse mapping pattern, random spreading model of sparse vectors, and transmission of spread sequence.

\begin{figure}[htbp]
\centerline{\includegraphics[width=0.45\textwidth]{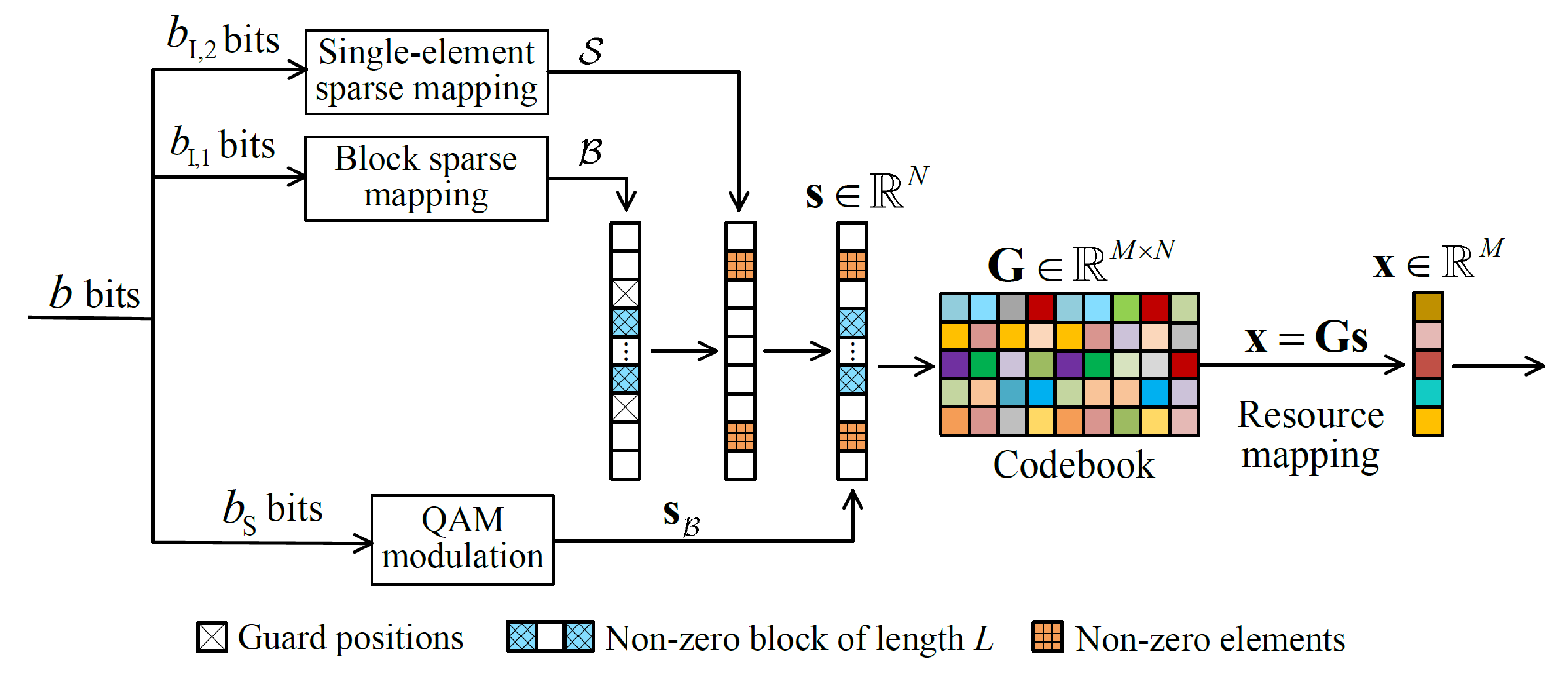}}  \vspace{-0pt}
\caption{Diagram of the sparse mapping process of DM-SVC. $b$ bits of information are mapped to a sparse vector of length $N$ containing $K_{\rm b}$ non-zero blocks and $K_{\rm s}$ non-zero elements.\vspace{-0pt}}
\label{fig1}
\end{figure}

\vspace{-0pt}
\subsection{Dual Sparse Mapping Pattern of DM-SVT}
Consider a short-packet communication system, where each data packet is assumed to contain $b$ bits of information. As shown in Fig. 1, the sparse mapping process of DM-SVC is divided into three steps. At the first step, $b_{\rm I, 1}$ bits are mapped to the $K_{\rm b}$ non-zero block index of a sparse vector ${\bf {s}}_{1}\in \mathbb{R}^{N}$ after block sparse mapping. At the second step, $b_{\rm I, 2}$ bits are mapped to the $K_{\rm s}$ non-zero index ${\cal S}$ of a sparse ${\bf {s}}_{1}\in \mathbb{R}^{N}$ after single-element sparse mapping. At the last step, the two sparse vectors are merged into one sparse vector, \emph{i.e.}, 
\begin{equation}
{\bf{s}} = \sqrt \alpha  {{\bf{s}}_1} + \sqrt {1 - \alpha } {{\bf{s}}_2},
\end{equation}
where $\alpha $ denotes the power allocation ratio of block sparse vector ${{\bf{s}}_1}$. The remaining $b_{\rm S}$ bits are then mapped to the values of $K_{\rm b}L+K_{\rm s}$ non-zero elements after QAM modulation, where $L$ denotes the length of non-zero blocks.  
In order to avoid the overlapping of single-element non-zero indexes and non-zero block indexes, the two positions before and after each non-zero block are used as guard positions. Therefore, there are a total of $2K_{\rm b}$ guard positions. After sparse mapping, a codebook matrix ${\bf{G}}$ is used to randomly spread the sparse vector $\bf s$. The resulting spread sequence $\bf x$ is then mapped to TF resources for transmission.

The block sparse mapping process involves selecting $K_{\rm b}$ non-zero blocks of length $L$ from $N$ positions, with a total number of ${\tbinom{N - {K_{\rm{b}}}(L - 1)}{{{K_{\rm{b}}}}}} $ possible combinations. Therefore, the amount of information bits that can be encoded by block sparse mapping is given by
\begin{equation}
{b_{{\rm{I,1}}}} = \left\lfloor {{{\log }_2}\left( {\begin{array}{*{20}{c}}
{N - {K_{\rm{b}}}(L - 1)}\\
{{K_{\rm{b}}}}
\end{array}} \right)} \right\rfloor .
\end{equation}

Considering the already selected ${{K_{\rm{b}}}L}$ non-zero block positions and their ${2{K_{\rm{b}}}}$ guard positions, there are ${N - {K_{\rm{b}}}L - 2{K_{\rm{b}}}}$ positions left for single-element sparse mapping. Thus, the number of information bits mapped to the indexes of ${{K_{\rm{s}}}}$ non-zero elements is given by
\begin{equation}
{b_{{\rm{I,2}}}} = \left\lfloor {{{\log }_2}\left( {\begin{array}{*{20}{c}}
{N - {K_{\rm{b}}}L - 2{K_{\rm{b}}}}\\
{{K_{\rm{s}}}}
\end{array}} \right)} \right\rfloor.
\end{equation}

In addition, the number of information bits mapped onto the ${K_{\rm{b}}}L + {K_{\rm{s}}}$ QAM symbol is
\begin{equation}
{b_{\rm{S}}} = ({K_{\rm{b}}}L + {K_{\rm{s}}}){\log _2}({M_{\bmod }}),
\end{equation}
where ${M_{\bmod }}$ is the modulation order.

\subsection{Random Spreading of DM-SVC}
After sparse mapping, the non-zero elements in the sparse vector are spread through $K$ pseudo-random codewords. This process can be described as
\begin{equation}
{\bf{x}} = \sqrt \alpha  \sum\nolimits_{k = {{\cal B}_1}}^{{{\cal B}_{L{K_{\rm{b}}}}}} {{{\bf{g}}_k}s_1^{(k)}}  + \sqrt {1 - \alpha } \sum\nolimits_{i = {{\cal S}_1},i \ne k}^{{{\cal S}_{{K_{\rm{s}}}}}} {{{\bf{g}}_i}s_2^{(i)}} , 
\end{equation}
where ${\bf{x}} \in {\mathbb{C}^M}$ denotes the transmitted TF-domain signal, ${{{\bf{g}}_{k}}}$ is the $k$-th codeword, ${s_1^{(k)}}$ and ${s_2^{(i)}}$ are the modulated symbols corresponding to the $k$-th non-zero element in ${\bf s}_{1}$ and the $i$-th non-zero element in ${\bf s}_{2}$, respectively. The codebook matrix can be expressed as ${\bf{G}} = [{{\bf{g}}_{1}},{{\bf{g}}_{2}},\cdots ,{{\bf{g}}_{N}}] \in {\mathbb{R}^{M\times N}}$. The elements of codebook $\bf G$ are composed of 1 and -1, both following a Bernoulli distribution \cite{Ji2018}. An example of $\bf G$ for $M=4$ and $N=5$ is given by
\begin{equation}
{\bf{G}} = \sqrt {\frac{1}{{{K_{\rm{b}}}L + {K_{\rm{s}}}}}} \left[ {\begin{array}{*{20}{c}}
{ - 1}&1&1&{ - 1}&1\\
1&{ - 1}&{ - 1}&1&{ - 1}\\
1&{ - 1}&1&{ - 1}&1
\end{array}} \right]
\end{equation}

\subsection{Transmission of Spread Sequence}
After performing inverse discrete Fourier transform (IDFT) on $\bf x$, the spread sequence is converted to time domain, \emph{i.e.}, ${{\bf{x}}_{\rm{T}}} = {{\bf{F}}^{\rm H}}{\bf{x}}$, where ${\bf{F}} \in {\mathbb{C}^{{M} \times {M}}}$ is a normalized DFT matrix with ${{\bf F}_{m,n}} = \frac{1}{{\sqrt M }}\exp ( - j2\pi mn/M)$. A $L_{\rm{CP}}$-length cyclic prefix (CP) is added to $\bf{x}_{\rm{T}}$ to avoid inter-symbol interference. At the receiver, after implementation of DFT on received signal and CP removal, the received signal can be expressed as
\begin{equation}
\begin{aligned}
{\bf{y}} &= {\bf{F}}{{\bf{H}}_{\rm{T}}}{{\bf{F}}^{\rm{H}}}{\bf{G}}{\bf{s}} + {\bf{w}}\\
 &= {{\bf{H}}_{\rm{F}}}{\bf{Gs}} + {\bf{w}}
\end{aligned},
\label{eq4}
\end{equation}
where ${{\bf{H}}_{\rm{T}}} \in {\mathbb{C}^{M \times M}}$ and ${{\bf{H}}_{\rm{F}}} \in {\mathbb{C}^{M \times M}}$ denote the time-domain and frequency-domain channel matrix, respectively. ${\bf{w}} \sim {\cal C}{\cal N}(0,{\sigma ^2}{{\bf{I}}_{N}})$ is the additive white Gaussian \mbox{noise (AWGN)} vector with variance $\sigma ^2$.

In static or low-speed moving scenarios, the time-domain channel gains remain constant within the duration of a transmission block. Therefore, the frequency-domain channel matrix ${{\bf{H}}_{\rm{F}}}$ can be regarded as a diagonal matrix \cite{Zhang2022,ZYFTWC25}.

At the receiver, our goal is to identify the ${{K_{\rm{b}}}L + {K_{\rm{s}}}}$ non-zero indexes of $\bf s$ from the received signal $\bf y$, so as to recover ${b_{{\rm{I,1}}}}+{b_{{\rm{I,2}}}}$ bits. In traditional SVC-type schemes, to ensure the recovery accuracy of $\bf s$, the number of TF resources (subcarriers) needs to satisfy \cite{Ji2018}  
\begin{equation}
{M^{{\rm{SVC}}}} = O\left( C{({K_{\rm{b}}}L + {K_{\rm{s}}})\log \left( {\frac{N}{{{K_{\rm{b}}}L + {K_{\rm{s}}}}}} \right)} \right).
\end{equation}
where $C$ is a positive constant \cite{Eldar2011}.

Considering that the $K_{\rm s}$ non-zero elements in ${\bf s}_2$ are regarded as non-zero blocks, there are a total number of $K_{\rm b}+K_{\rm s}$ non-zero blocks in $\bf s$. According to the block sparse recovery theory \cite{Eldar2011}, the number of subcarriers required by the MD-SVC scheme needs to satisfy
\begin{equation}
{M^{{\rm{MD-SVC}}}} = O\left(C {({K_{\rm{b}}} + {K_{\rm{s}}})\log \left( {\frac{N}{{{K_{\rm{b}}}L + {K_{\rm{s}}}}}} \right)} \right).
\end{equation}

\begin{figure}[htbp]
\centerline{\includegraphics[width=0.35\textwidth]{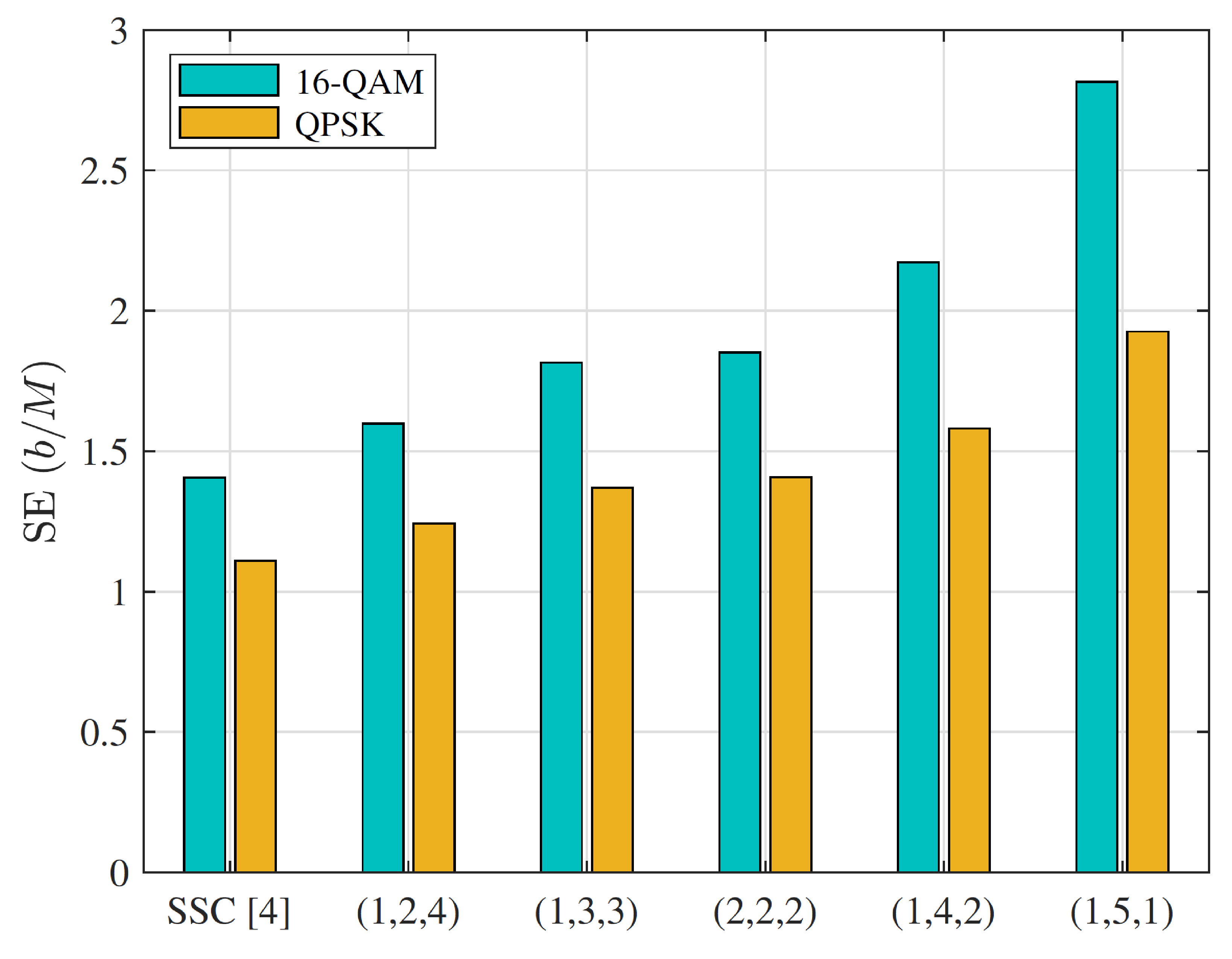}}\vspace{-0pt}
\caption{Comparison of SE of the SSC scheme \cite{ZhangXuewan2022} and DM-SVC scheme with different parameters $({{K_{\rm{b}}}},L,{K_{\rm{s}}})$, where $N=138$ and $C=5$. }
\label{fig2}
\vspace{-0pt}
\end{figure}

The SE of DM-SVC scheme can be expressed as
\begin{equation}
{\rm{S}}{{\rm{E}}^{{\rm{MD - SVC}}}} = \frac{{{b_{{\rm{I,1}}}} + {b_{{\rm{I,2}}}} + {b_{\rm{S}}}}}{{{M^{{\rm{MD - SVC}}}}}},
\end{equation}
where ${b_{\rm{S}}} = ({K_{\rm{b}}}L + {K_{\rm{s}}}){\log _2}({M_{\bmod }})$. Note that the SE of the conventional SSC scheme \cite{ZhangXuewan2022} is given as
\begin{equation}
{\rm{S}}{{\rm{E}}^{{\rm{ESVC}}}} = \frac{{{b_{\rm{I}}} + {b_{\rm{S}}}}}{{{M^{{\rm{SVC}}}}}},
\end{equation}
where ${{b_{\rm{I}}}}={\rm{log}_2}{\tbinom{N}{{{K_{\rm{b}}L+K_{\rm s}}}}} $ denotes the number of bits encoded into the non-zero indexes in the SSC scheme \cite{ZhangXuewan2022}.

By comparing (10) and (11), it can be observed that the number of bits encoded by the MD-SVC scheme is less than that of the SSC scheme \cite{ZhangXuewan2022} because ${b_{{\rm{I,1}}}} + {b_{{\rm{I,2}}}} < {b_{\rm{I}}}$. However, the SSC scheme requires more subcarriers to ensure the transmission reliability as compared to the DM-SVC scheme. Fig. 2 shows the comparison between SSC scheme \cite{ZhangXuewan2022} and DM-SVC scheme in terms of SE. The total number of non-zero elements in both schemes is set to 6. It can be observed that the SE of the DM-SVC scheme with different parameters is higher than that of the SSC scheme. The DM-SVC scheme achieves more than 12\% and 14\% SE improvement compared to the SSC scheme for QPSK and 16-QAM modulation, respectively. When the parameter $({{K_{\rm{b}}}},L,{K_{\rm{s}}}) =(1, 5,1)$ is adopted, the DM-SVC achieves a two-fold SE improvement compared to the SSC scheme for 16-QAM modulation.

%\addtolength{\topmargin}{-0.87cm}

\vspace{-0pt}
\section{DM-SVC Decoding}
In this section, the decoding algorithm for DM-SVC scheme is introduced. The goal of DM-SVC decoding is to recover the non-zero support as well as the non-zero value of the sparse vector $\bf s$ from the received signal $\bf y$. The $b$ bits of information is then recovered through inverse sparse mapping and symbol demodulation.

Note that the single-element sparsity $K_{\rm s}$, block sparsity $K_{\rm b}$ and block length $L$ are known to both transmitter and receiver, the maximum likelihood (ML) detection problem for the system model in (7) is
\begin{equation}
{{\bf{s}}^*} = \mathop {\arg \max }\limits_{||{\bf{s}}|{|_0} = {K_{\rm{b}}}L + {K_{\rm{s}}}} \Pr ({\bf{y}}|{\bf{s}},{\bf{\Phi }}),
\end{equation}
where $||{\bf{s}}|{|_0}$ is the ${\ell _0}$-norm of $\bf s$ counting the number of non-zero elements in $\bf s$, ${\bf{\Phi }}{\rm{ = }}{{\bf{H}}_{\rm{F}}}{\bf{G}} \in {\mathbb{C}^{M \times N}}$ is the measurement matrix. Since our goal is to find out the support of $\bf s$, we alternatively have
\begin{equation}
\begin{aligned}
&\min ||{\bf{y}} - {\bf{\Phi }}(\sqrt \alpha  {[{{\bf{s}}_1}]_{\cal B}} + \sqrt {1 - \alpha } {[{{\bf{s}}_2}]_{\cal S}})|{|^2}\\
&{\rm{s}}.{\rm{t}}.\;{\rm{   }}{\kern 1pt} |{\cal B}| = {K_{\rm{b}}}L,|{\cal S}| = {K_{\rm{s}}}
\end{aligned}.
\label{eq13}
\end{equation}
To find the ML solution, we need to enumerate all possible candidate support combinations with cardinality ${K_{\rm{b}}}L + {K_{\rm{s}}}$. Unfortunately, this exhaustive search would not be feasible in most practical situations. A standard compressed sensing model is shown in (13). The existing MMP \cite{Kwon2014} can be used to solve the sparse recovery problem in (13). However, the MMP algorithm only considers the prior information about the number of non-zero elements in the sparse vector, without exploiting the underlying structured sparsity information (\emph{e.g.}, the non-zero block information in $\bf s$), which limits its decoding reliability in the model shown in (13). To address this issue, a two-stage decoding algorithm is proposed to sequentially identify $K_{\rm b}$ non-zero block indexes and $K_{\rm s}$ single-element non-zero indexes.

\renewcommand{\algorithmicrequire}{\textbf{Input:}}
\renewcommand{\algorithmicensure}{\textbf{Output:}}
\begin{algorithm}[h]
\caption{\footnotesize{\bf {Algorithm 1~}}Two-Stage Decoding for DM-SVC Scheme }
  \begin{algorithmic}[1]
  \footnotesize  \Require 
received signal ${\bf{y}}$, measurement matrix $\bf \Phi$, power allocation ratio $\alpha$, number of non-zero blocks $K_{\rm b}$, number of non-zero elements $K_{\rm s}$, and non-zero block length $L$.
\State Initialization: ${\bf r}^{(0)} = \bf y$, ${{\cal T}^{(0)}} = \emptyset $, ${{\cal B}^{(0)}} = \emptyset $.
\State \textbf{Stage I}: Identify the indexes of $K_{\rm b}$ non-zero blocks.
\For {$k=1,2,\cdots,K_{\rm b}$}
\For {$l=0,2,\cdots,L-1$}
\State ${{\bf{U}}^{(k,l)}} = {\bf{\Phi }}{{\bf{\Pi }}^{-l}},\quad l=0,1,\cdots,L-1$
\EndFor
\State $\{ q_k^*,{l^*}\}  = \mathop {\arg \max }\limits_{q = 1,2, \cdots ,B;l = 0,1, \cdots ,L-1} ||{({\bf{U}}_q^{(k,l)})^{\rm{H}}}{{\bf{r}}_k}||_2^2$.
\State Find out the block index $q_k^*$ and the number of cyclic\\  \quad\quad shifts ${l^*}$ according to (19).
\State Obtain the support set ${{\cal T}^{(k)}}$ according to (20).
\State Estimate non-zero value according to (21).
\State Update residual according to (22).
\State Perform symbol demodulation according to (23).
\EndFor
\State Merge support sets according to (24).
\State Interference cancellation: ${{\bf{y}}_{\rm{s}}} = {\bf{y}} - \sqrt \alpha  {{\bf{\Phi }}_{\hat {\cal B}}}{[{{{\bf{\hat s}}}_1}]_{\hat {\cal B}}}$
\State \textbf{Stage II}: Identify the indexes of $K_{\rm s}$ non-zero elements.
\State Obtain the support set ${{\cal S}^*}$ by using the MMP algorithm.
\State Estimate the $K_{\rm s}$ non-zero values: ${[{{{\bf{\hat s}}}_2}]_{{{\cal S}^*}}} = \frac{1}{{\sqrt {1 - \alpha } }}{\bf{\Phi }}_{{{\cal S}^*}}^\dag {{\bf{y}}_{\rm{s}}}$.
    \Ensure Index sets ${\cal B}{^*}$ and ${\cal S}{^*}$, and non-zero values ${[{{{\bf{\hat s}}}_1}]_{ {\cal B}^{*}}}$ and ${[{{{\bf{\hat s}}}_2}]_{ {\cal S}^{*}}}$.      
  \end{algorithmic}
\end{algorithm}
\vspace{-0pt}
%\addtolength{\topmargin}{-0.02in}

%\addtolength{\topmargin}{-0.395in}
\vspace{-0pt}
\subsection{Stage I: Identification of $K_{\rm b}$ Non-Zero Block Indexes}
Note that the signal model in (7) can be rewritten as
\begin{equation}
\begin{aligned}
{\bf{y}} &= \sqrt \alpha  {\bf{\Phi }}{{\bf{s}}_1} + \sqrt {1 - \alpha } {\bf{\Phi }}{{\bf{s}}_2} + {\bf{w}}\\
 &= \sqrt \alpha  {\bf{\Phi s}} + {\bf{\tilde w}}
\end{aligned},
\end{equation}
where ${\bf{\tilde w}} = \sqrt {1 - \alpha } {\bf{\Phi }}{{\bf{s}}_2} + {\bf{w}}$ is an equivalent noise signal. The support set of non-zero blocks can be obtained by solving the following optimization problem
\begin{equation}
{{\cal B}^*} = \mathop {\arg \min }\limits_{|{\cal B}| = {K_{\rm{b}}}L} ||{\bf{y}} - \sqrt \alpha  {\bf{\Phi }}{[{{\bf{s}}_1}]_{\cal B}})|{|_2^2}.
\label{15}
\end{equation}

%\addtolength{\topmargin}{0.15in}

Due to the nonuniform distribution of $K_{\rm b}$ non-zero blocks in ${\bf s}_1$, the traditional block orthogonal matching pursuit algorithm cannot be directly used to solve (15). By utilizing the circular shift property of matrix-vector operations, one can obtain
\begin{equation}
{\bf{\Phi s}} = {\bf{\Phi }}{{\bf{\Pi }}^l}{{\bf{\Pi }}^{ - l}}{\bf{s}},
\end{equation}
where ${\bf{\Pi }} \in {\mathbb{R}^{N \times N}}$ is a permutation matrix, \emph{i.e.},
\begin{equation}
{\bf{\Pi }} = {\rm{circ}}\{ [0,1,0, \cdots ]_N^{\rm{T}}\} .
\end{equation}
Note that ${{\bf{\Pi }}^0} = {{\bf{I}}_N}$ and ${{\bf{\Pi }}^{ - l}}{{\bf{\Pi }}^l} = {{\bf{I}}_N}$ are two basic properties of permutation matrix.

To find the index of the $k$-th nonzero block, the columns of the measurement matrix are first cyclically shifted to the left $l$ times at the $k$-th iteration, \emph{i.e.},
\begin{equation}
{{\bf{U}}^{(k,l)}} = {\bf{\Phi }}{{\bf{\Pi }}^{-l}},\quad l=0,1,\cdots,L-1.
\end{equation}
Then, the matrix ${{\bf{U}}^{(k,l)}}$ is divided into $B$ sub-matrices and the block correlation values between each sub-matrix ${\bf{U}}_q^{(k,l)}$ and residual are calculated. The non-zero block index that maximizes the block correlation value and the number of cyclic shifts is given as
\begin{equation}
\{ q_k^*,{l^*}\}  = \mathop {\arg \max }\limits_{q = 1,2, \cdots ,B;l = 0,1, \cdots ,L-1} ||{({\bf{U}}_q^{(k,l)})^{\rm{H}}}{{\bf{r}}^{(k)}}||_2^2,
\end{equation}
where $B = N/L$ is an integer, ${{\bf{r}}_k}$ is the residual updated for the $k$-th iteration. When $k=0$, ${{\bf{r}}^{(0)}} = {\bf{y}}$. The support set of the $k$-th non-zero block can be obtained as

\begin{equation}
{{\cal T}^{(k)}} = \{ q_k^*(B - 1) + {l^*} + 1, \cdots ,q_k^*B + {l^*}\}.
\end{equation}
${{\cal T}^{(k)}}$ is then merged into ${{\cal B}^{(k)}}$, \emph{i.e.}, ${{\cal B}^{(k)}} = {{\cal B}^{(k)}} \cup {{\cal T}^{(k)}}$. The non-zero value of the $k$-th block is estimated as
\begin{equation}
{[{{{\bf{\hat s}}}_1}]_{{{\cal T}^{(k)}}}} = \frac{1}{{\sqrt \alpha  }}{\bf{\Phi }}_{{\cal T}^{(k)}}^\dag {{\bf{r}}^{(k)}}.
\end{equation}

The residual at the $(k+1)$-th iteration is then updated by
\begin{equation}
{{\bf{r}}^{(k+1)}} = {\bf{y}} - \sqrt \alpha  {{\bf{\Phi }}_{{{\cal B}^{(k)}}}}{[{{{\bf{\hat s}}}_1}]_{{{\cal B}^{(k)}}}}
\end{equation}

Based on the estimated non-zero values, the $l$-th symbol modulated on the $k$-th non-zero block can be estimated as
\begin{equation}
\tilde s_1^{(l)} = \mathop {\arg \min }\limits_{{a_m} \in {\cal A},\forall m \in \{ 1,2, \cdots ,{M_{\bmod }}\} } |{[{{{\bf{\hat s}}}_1}]_{{{\cal T}^{(k)}_l}}} - {a_m}|,
\end{equation}
where ${\cal A} = \{ {a_1},{a_2}, \cdots ,{a_{{M_{\bmod }}}}\} $ is a QAM modulation alphabet with size $M_{\bmod }$.

The support set of ${\bf s}_{1}$ can be estimated by merging the $K_{\rm b}$ support sets
\begin{equation}
{\cal B}^{*} = {{\cal T}^{(1)}} \cup {{\cal T}^{(2)}} \cup  \cdots  \cup {{\cal T}^{({K_{\rm{b}}})}}.
\end{equation}
Finally, ${b_{{\rm{I,1}}}}$ bits of information mapped on non-zero block indexes can be obtained by block sparse demapping.

\vspace{-0pt}
\subsection{Stage II: Identification of $K_{\rm s}$ Non-Zero Indexes}

The goal of Stage II is to recover the $K_{\rm s}$ single-element non-zero indexes and their corresponding values. With the estimated non-zero block support in hand, the received signal corresponding to ${\bf s}_{2}$ can be expressed as
\begin{equation}
{{\bf{y}}_{\rm{s}}} = {\bf{y}} - \sqrt \alpha  {{\bf{\Phi }}_{{\cal B}^{*}}}{[{{{\bf{\hat s}}}_1}]_{ {\cal B}^{*}}},
\end{equation}
where ${[{{{\bf{\hat s}}}_1}]_{ {\cal B}^{*}}} = {[{({[{{{\bf{\hat s}}}_1}]_{{{\cal T}^{(1)}}}})^{\rm{H}}},{({[{{{\bf{\hat s}}}_1}]_{{{\cal T}^{(2)}}}})^{\rm{H}}}, \cdots ,{({[{{{\bf{\hat s}}}_1}]_{{{\cal T}^{({K_{\rm{b}}})}}}})^{\rm{H}}}]^{\rm{H}}}$.

The support set of ${\bf s}_{2}$ can be obtained by solving the following optimization problem
\begin{equation}
{{\cal S}^*} = \mathop {\arg \min }\limits_{|{\cal S}| = {K_{\rm{s}}}} ||{{\bf{y}}_{\rm{s}}} - \sqrt {1 - \alpha } {\bf{\Phi }}{[{{\bf{s}}_2}]_{\cal S}}||_2^2.
\end{equation}
In this paper, the existing MMP \cite{Kwon2014} is adopted to solve the sparse recovery problem in (26). The estimated non-zero indexes are then used to recover the ${b_{{\rm{I,2}}}}$ bits of information by sparse demapping. The values of the $K_{\rm s}$ non-zero elements can be estimated as
\begin{equation}
{[{{{\bf{\hat s}}}_2}]_{{{\cal S}^*}}} = \frac{1}{{\sqrt {1 - \alpha } }}{\bf{\Phi }}_{{{\cal S}^*}}^\dag {{\bf{y}}_{\rm{s}}}.
\end{equation}

Finally, the symbols carried on the $K_{\rm s}$ non-zero elements are estimated in a manner similar to that in (23).

The pseudo-code of proposed two-stage decoding algorithm is shown in \textbf{Algorithm 1}. It is noteworthy that the proposed two-stage decoding algorithm exploits block-structured sparsity in the Stage I to enhance the identification accuracy of non-zero block indexes, thus achieving more reliable decoding performance than the traditional MMP algorithm. The effectiveness of proposed two-stage decoding algorithm will be verified in simulation results in Section V.

%\addtolength{\topmargin}{0.04in}
  \vspace{-0pt}
\section{Complexity Analysis}
As shown in \textbf{Algorithm 1}, the decoding complexity of DM-SVC mainly lies in the detection of $K_{\rm b}$ non-zero block indexes and $K_{\rm s}$ non-zero indexes.
At Stage I, the complexity of calculating block correlation and estimating non-zero values is given as $O((LMN + M{L^2} + {L^3}){K_{\rm{b}}})$. At Stage II, the MMP algorithm is used to estimate the remaining $K_{\rm s}$ non-zero indexes with a complexity of $O(K_{\rm{s}}^3 + MK_{\rm{s}}^2 + {L_{\rm{p}}}{K_{\rm{s}}}(MN + 2M))$. Considering that $L$ and ${K_{\rm{s}}}$ are usually much smaller than $M$, the total complexity of proposed two-stage decoding algorithm can be summarized as $O((LMN + M{L^2}){K_{\rm{b}}} + MK_{\rm{s}}^2 + {L_{\rm{p}}}{K_{\rm{s}}}(MN + 2M))$, where $L_{\rm p}$ is the maximum searching paths in the MMP algorithm \cite{Kwon2014}. The complexity of the MMP decoding algorithm used in the SSC scheme \cite{ZhangXuewan2022} is given as $O(M{{\tilde K}^2} + {L_{\rm{p}}}\tilde K(MN + 2M)$, where ${\tilde K}$ is the number of non-zero elements in SSC scheme. Since $K_{\rm b}$ and $L$ usually take smaller values, the proposed two-stage decoding algorithm and the MMP decoding algorithm used in \cite{ZhangXuewan2022} have comparable complexity when $\tilde K = {K_{\rm{b}}}L + {K_{\rm{s}}}$.

%\addtolength{\topmargin}{0.02in}

\section{Simulation Results}

The BLER performance and power allocation ratio of proposed DM-SVC scheme are evaluated in this section. The existing SVC \cite{Ji2018}, ESVC \cite{Kim20202}, SSC \cite{ZhangXuewan2022}, GSPARC \cite{Sinha2024TCOM} and BOSS \cite{DhanTWC2023} schemes are selected for performance comparison. QPSK modulation is adopted for ESVC, SSC, GSPARC, BOSS and DM-SVC schemes. For all comparison schemes, the number of non-zero elements in sparse vectors is the same. The BLER is defined as ratio of number of packets received in error to total number of transmitted packets.

\begin{figure}[htbp]
\centerline{\includegraphics[width=0.38\textwidth]{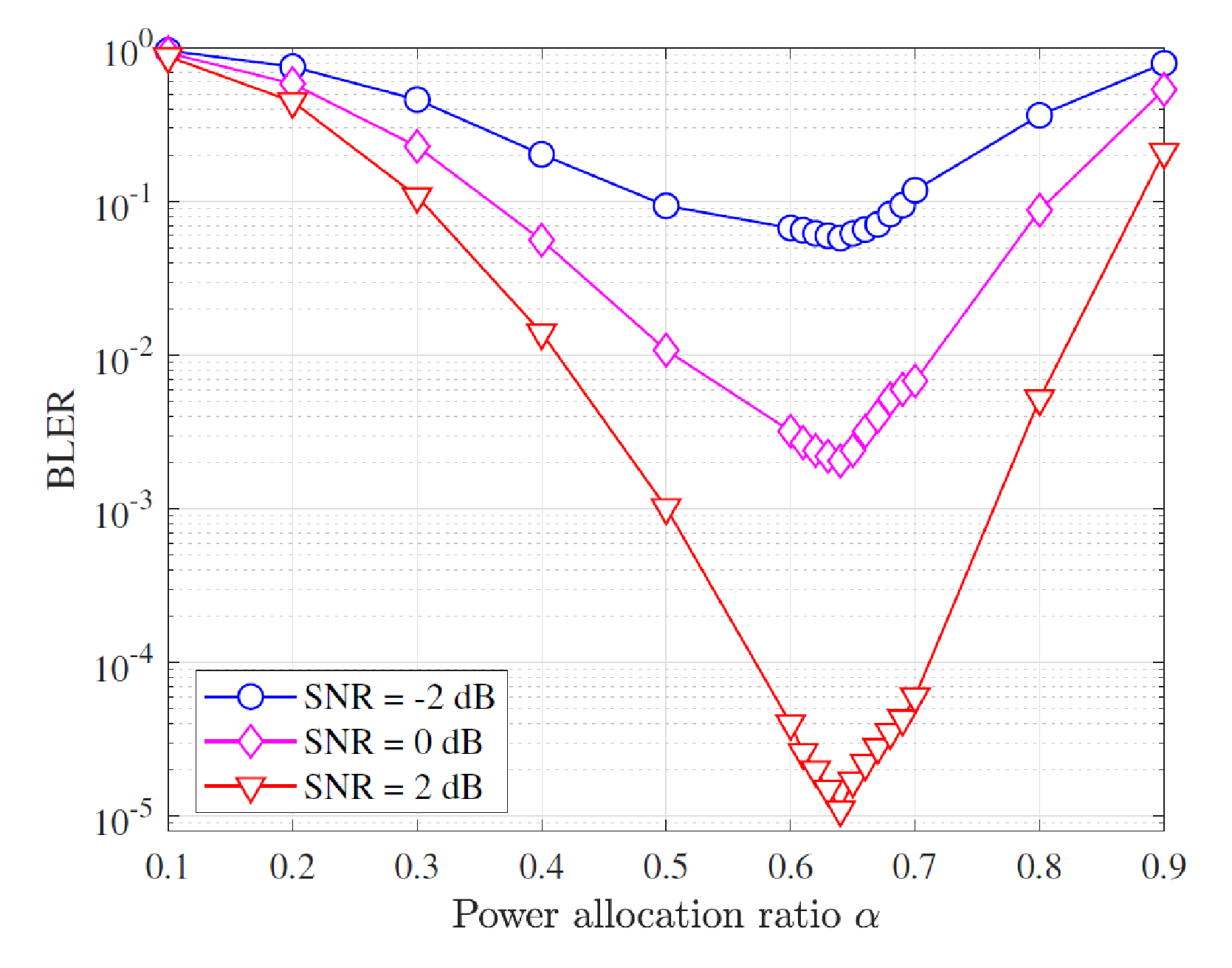}}\vspace{-0pt}
\caption{BLER vs. power allocation ratio in DM-SVC scheme with $K_{\rm b}=1$, $L=3$, $K_{\rm s}=1$, $N=2100$, $M=96$ and $b=30$. \vspace{-0pt}}
\label{fig3}
\end{figure}

\begin{figure}[htbp]
\centerline{\includegraphics[width=0.36\textwidth]{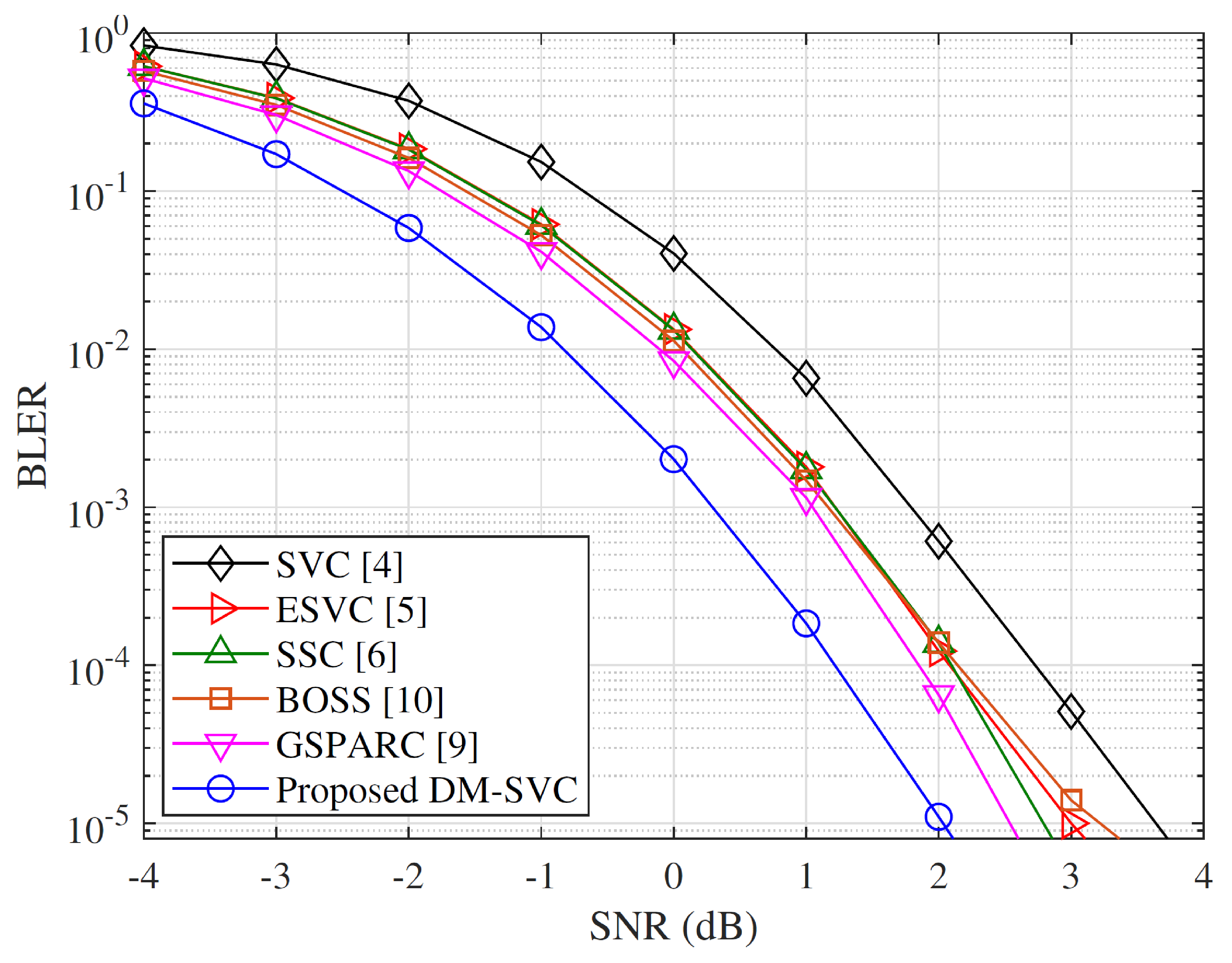}}\vspace{-0pt}
\caption{BLER of different schemes vs. SNR with $\alpha=0.64$, $K_{\rm b}=1$, $L=3$, $K_{\rm s}=1$, $N=2100$, $M=80$ and $b=30$. \vspace{-0pt}}
\label{fig4}
\end{figure}

Fig. 3 shows the BLER of proposed DM-SVC scheme as a function of power allocation ratio $\alpha$. The number of non-zero blocks is kept consistent. It can be observed that the optimal power allocation ratio at different SNR levels is approximately $\alpha^{*}=0.64$. This indicates that the identification of non-zero block indexes at the first stage plays a dominant role in decoding performance. This is because at the first stage more power needs to be allocated to non-zero blocks to combat interference from single-element non-zero elements. At the second stage, the interference from non-zero blocks have been removed before identifying non-zero element indexes, thus requiring less power.

Fig. 4 shows the BLER of proposed DM-SVC scheme in comparison to the existing schemes. The SE of all comparison schemes is fixed at SE$=0.375$ bps/Hz. Regarding the BOSS scheme, the codebook matrix consists of 64 mutually orthogonal Walsh-Hadamard matrices \cite{DhanTWC2023}, each of size $80\times 80$. It can be seen that the proposed DM-SVC scheme outperforms the existing SVC, ESVC, SSC, GSPARC and BOSS schemes in terms of BLER. For example, at the target BLER$=10^{-5}$, an SNR gain of about 0.5 dB can be achieved by the proposed DM-SVC scheme over the GSPARC scheme.

Fig. 5 investigates the impact of the length of non-zero blocks $L$ and number of non-zero elements $K_{\rm s}$ on the BLER performance of DM-SVC scheme. It can be observed that the reliability decreases as the $L$ and $K_{\rm s}$, and consequently the number of transmitted bits $b$ also increases. The DM-SVC scheme provides diverse choices for short-packet communications by controlling the parameters $L$ and $K_{\rm s}$. In practical communications, the parameters $L$ and $K_{\rm s}$ can be flexibly selected according to the requirements of SNR, reliability, number of transmitted bits, and SE.

\begin{figure}[htbp]
\centerline{\includegraphics[width=0.38\textwidth]{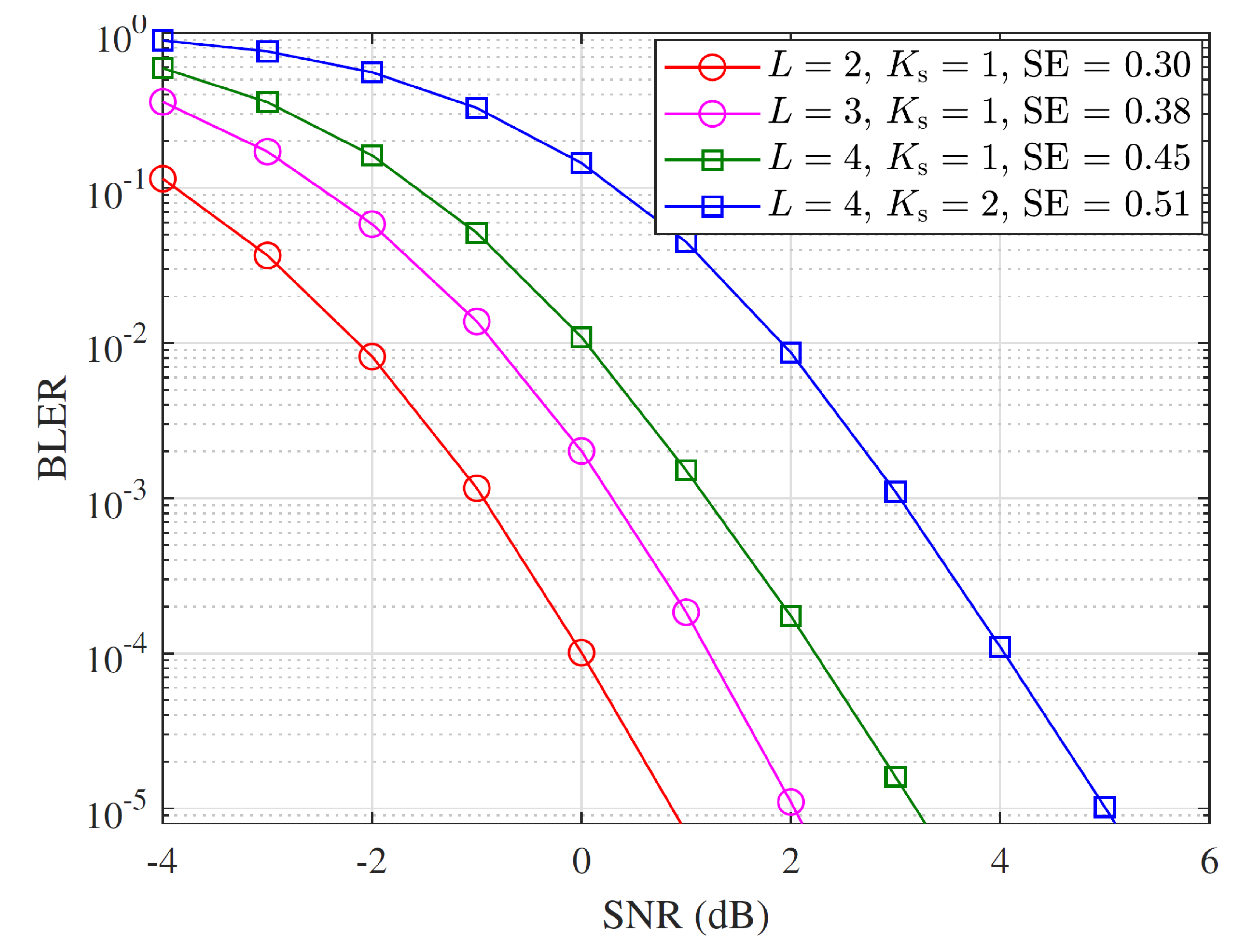}}\vspace{-0pt}
\caption{Impact of $L$ and $K_{\rm s}$ on BLER performance in proposed DM-SVC scheme with $K_{\rm b}=1$ and $M=80$. \vspace{-0pt}}
\label{fig5}
\end{figure}

\begin{figure}[htbp]
\centerline{\includegraphics[width=0.38\textwidth]{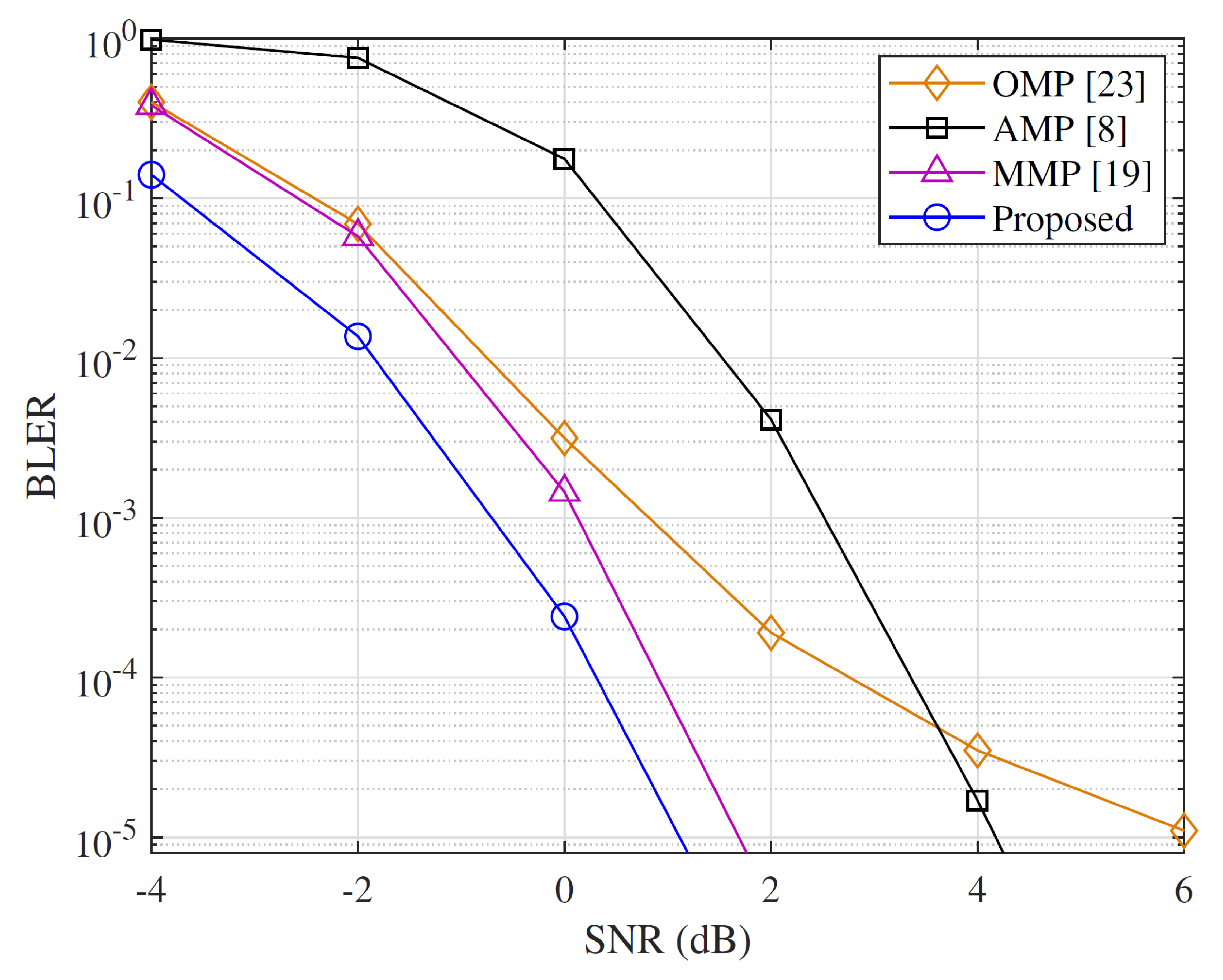}}\vspace{-0pt}
\caption{BLER of different algorithms vs. SNR in DM-SVC scheme with $K_{\rm b}=1$, $L=2$, $K_{\rm s}=1$, $N=262$, $M=80$ and $\alpha = 0.64$. \vspace{-0pt}}
\label{fig6}
\vspace{-0pt}
\end{figure}

Fig. 6 investigates the BLER performance of different decoding algorithms in DM-SVC scheme. It can be seen that the proposed two stage algorithm achieves better BLER performance compared to the existing AMP \cite{CRushTIT2021}, MMP \cite{Kwon2014} and OMP \cite{OMP2020} algorithms. The superior performance of proposed decoding algorithm is mainly attributed to its exploitation of block-structured sparsity at the first stage, which enhances the identification accuracy of non-zero block indexes.

\vspace{-0pt}
\section{Conclusions}
In this paper, a DM-SVC-based short-packet transmission scheme is proposed for URLLC. The proposed DM-SVC scheme introduces two sparse mapping patterns to enhance the transmission performance. The block-structured sparse prior information introduced by block sparse mapping pattern is used to design a high-accuracy two-stage decoding algorithm. Simulation results show that the proposed DM-SVC scheme outperforms the existing SVC-type schemes in terms of SE and reliability. 
%In the future, the proposed DM-SVC will be extended to massive MIMO systems, and low-complexity decoding algorithms will be further investigated.

\vspace{-0pt}
\bibliographystyle{IEEEbib}
\bibliography{refs}

\end{document}